# DNA detection by THz pumping

Andrey L. Chernev[1], Nicolay T. Bagraev[2], Leonid E. Klyachkin[2], Anton K. Emelyanov[1], Michael V. Dubina[1]

[1] St. Petersburg Academic University – Nanotechnology Research and Education Centre RAS, 8/3 Khlopin str., St. Petersburg 194021, Russia

[2] Ioffe Physical-Technical Institute of the Russian Academy of Sciences, 26 Polytechnicheskaya str., St. Petersburg 194021, Russia

**Prompt detection of oligonucleotides with semiconductor devices is a technique that is believed to be capable of improving all current genetics technologies[1]. Almost every detection method requires the use of fluorescent dyes and markers[2-5] for the indirect measurements of the nucleic acids' characteristics. The development of pH sensing has provided a significant improvement to the field of label-free, real-time, and non-optical semiconductor sequencing[6] and amplification[7]. Another promising conception is single-molecule nanopore analysis[8-10]. However, a superior method for oligonucleotide detection has yet to be developed. Here, we demonstrate that synthetic oligonucleotides can be directly detected without labels by their self-resonant modes with silicon nanosandwich pump device. The self-resonant modes of oligonucleotides are identified not only by Raman spectroscopy[11,12], but also with a silicon nanosandwich-based pump device that provides both the excitation of the oligonucleotides' self-resonant modes and feedback for current-voltage measurements. Our results demonstrate a new method for label-free, real-time oligonucleotide characterisation by their self-resonant modes, which are unique to their conformation and sequence. We anticipate that our assay will be used as a starting point for a more detailed investigation of the aforementioned mechanism, which can be used as a basis for oligonucleotide detection and analysis. Furthermore, this technique can be applied to improve existing modern genetics technologies.**

The real-time amplification and detection of nucleic acids has given rise to the development of life science research and molecular diagnostics[2,3,5]. These methods are now a basis of the techniques applied for the express detection and quantification of small amounts of nucleic acids and have a wide array of applications[2-5]. However, use of these techniques for the real-time detection of nucleic acids requires precision optics as well as fluorescently labelled, sequence-specific probes or fluorescent dyes for DNA labelling[3,5]. These requirements represent a significant disadvantage of such techniques because of the need to collect oligonucleotide signals indirectly. Several attempts have been made to resolve this issue. Recently, a semiconductor-based nucleic acid sequencer that uses the pH-sensing capability of ion-sensitive field-effect transistors (ISFET) has been demonstrated[6]. Another device that is capable of amplifying and detecting DNA simultaneously using embedded heaters, temperature sensors, and ISFET sensor arrays also appears to be highly effective[7]. The most important result of the studies mentioned was to simultaneously provide amplification and detection. Nevertheless, despite the development of ISFET technology[13-16], there are still challenges that it cannot address. The most crucial disadvantage of ISFET-based sensors is that they are based on a pH-sensing mechanism that is not target specific.

Here, we present a new method of oligonucleotide detection by the excitation of their self-resonant modes with silicon nanosandwich pump device, which correspond to the unique combination of the nucleotide sequence and entire molecular shape. This proposed oligonucleotide detection method is based on the interaction of a silicon nanosandwich with nucleic acids deposited on its surface. This silicon nanosandwich represents an ultra-narrow, p-type silicon quantum well (Si-QW) confined by delta barriers heavily doped with boron on the n-Si (100) wafer (Fig. 1a). The edge channels of this Si-QW have been shown to be an effective source of THz emission caused by the presence of the negative-U dipole boron centres (Figs. 1b, 1c and 1d)[17] (see the Methods Summary for details). To increase the selective THz line emission, the corresponding system of microcavities is introduced into the Si-QW plane[17]. Such devices allow for the creation of THz spectra that are similar to the self-resonant modes of oligonucleotides[11,12,18]. Thus, the excitation of the self-resonant modes of the oligonucleotides deposited in the Si-QW plane becomes possible and provides feedback that gives rise to the changes in the conductance of the edge channels.

The $U$-$I$ characteristics of the silicon nanosandwich prepared within the frameworks of the Hall geometry were measured under the stabilisation of a drain-source current to define the resonant frequencies of the oligonucleotides (Fig. 1a) (see the Methods Summary for details). The oligonucleotides were precisely deposited onto the delta barrier above the edge channels of the silicon nanosandwich with a micropipette and microfluidic system (Fig. 1e, 1f and 1g). The concentration was selected to provide no more than one oligonucleotide per microcavity (see the Methods Summary for details).

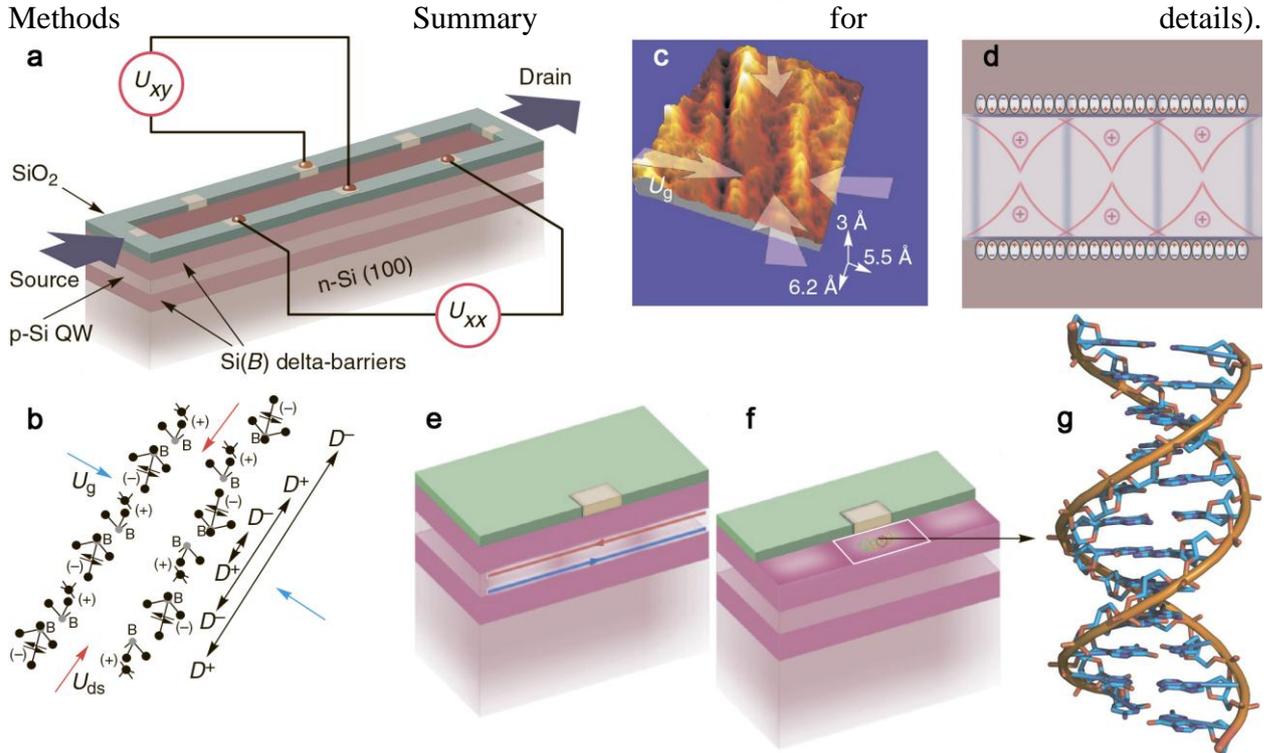

**Figure 1 | Scheme of the silicon nanosandwich. a,** The silicon nanosandwich representing the ultra-narrow p-type silicon quantum well confined by delta barriers heavily doped with boron on the n-type Si (100) wafer. **b,** The reconstructed trigonal dipole boron centres that result from the negative-U reaction, $2B^0 \rightarrow B^+ + B^-$. **c,** STM image of the upper delta barrier heavily doped with boron that demonstrates the chains of dipole boron centres oriented along the [011] axis. **d,** Diagram showing the wave packs of holes in the edge channels of the dipole boron centres in the delta barriers. **e,** Image of the contact region showing the edge channels scheme. **f, g,** Scheme of an oligonucleotide molecule placed on the edge channels confined by the dipole boron centres.

The changes of the longitudinal voltage $U_{xx}$ and lateral (Hall) voltage $U_{xy}$ appeared to demonstrate the resonant behaviour as a function of the $I_{ds}$ value (Figs. 2a, 2b). The $I_{ds}$ steps revealed by measuring the $U_{xx}$-$I_{ds}$ characteristics allow us to define the generation frequency ($f$) using the relation $I=epf$, where $I$ is the $I_{ds}$ value corresponding to a step on the $U_{xx}$-$I_{ds}$ characteristics and $p$ is the number of holes in the edge channels, which depends on the sheet density (see the Methods Summary section for details). This approach is given by the quantum pump operational method because the nanowire-turnstile device has been shown to perform as a quantum pump[19]. We use the same method in this study, with the only difference being that the nanowire with modulated barriers created by gating is replaced by the edge channels of the silicon nanosandwich (Fig. 1e). Moreover, the dipole boron centres in the edge channels have been shown to be magnetically ordered by the exchange interaction through the 2D holes[17]. Thus, the fragments of edge channels with single holes appear to be represented as independent quantum pumps, thereby providing the relation $I=epf$ for a large number of holes. Besides, the number of holes in the edge channels must define the parameters of the microcavities introduced into the Si-QW plane to only provide a single hole in each microcavity. Thus, by varying the sheet density of holes and the concentration of the oligonucleotides, a number of versions were realised to involve a single hole and single oligonucleotide in one microcavity. If the number of holes in the edge channel determined by $p_{2D}$ is considered (see the Methods Summary for details), then $p=120$, and the self-mode frequency $f$ is determined to be equal to 2.2 THz from the resonant value of the drain-source current of 43.8 µA (Fig. 2a).

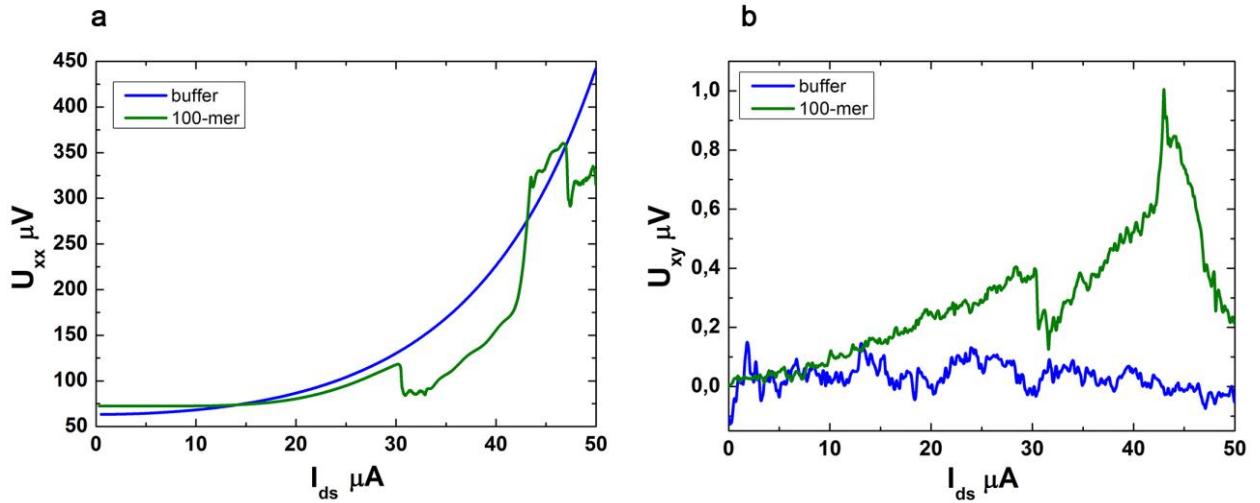

Figure 2| $U_{xx}$-$I_{ds}$ and $U_{xy}$-$I_{ds}$ characteristics. The $U_{xx}$-$I_{ds}$ (a) and $U_{xy}$-$I_{ds}$ (b) characteristics of the silicon nanosandwich without (blue) and with (green) 100-mer oligonucleotides. Measurements were performed at room temperature under the stabilization of the drain-source current.

In contrast, the $I_{ds}$ steps can be interpreted within the framework of the Josephson light emitting diode model[20]. Specifically, the edge channel can be presented as the subsequence of tunnel junctions formed by the negative-U dipole boron centres, with the optimal gain and generation obtained using the matching relation of sheet density with the confinement specified by the formation of microcavities in the Si-QW plane. Thus, the resonant frequencies caused by the tunnelling holes through the dipole boron centres and the oligonucleotide self-mode revealed by the feedback mechanism appear to result from the well-known relation[21] $f=2e\Delta U_{xx}/h$. Among the three resonant frequencies shown in Fig. 2A, the side features that demonstrate the negative differential resistance, i.e., 2.7 and 7.9 THz, are related to the Rabi splitting induced by the strong coupling between the tunnelling junction and microcavity modes. In contrast, the central positive differential resistance response indicates an oligonucleotide self-mode frequency of 2.8 THz, as revealed by the feedback mechanism. Thus, the two different approaches for the analysis of the $U_{xx}$-$I_{ds}$ measurements are found to be in good agreement.

The same model is used for the analysis of both the $U_{xx}$-$I_{ds}$ and $U_{xy}$-$I_{ds}$ characteristics because the conductance of the silicon nanosandwich is caused by the formation of the edge channels. Therefore, the $U_{xy}$-$I_{ds}$ measurements are described by the relation $f=2e\Delta U_{xy}/h$, which results from the counter fluxes of holes near the upper and lower delta barriers (Fig. 1d and 1e). This assumption is related to models of spin Hall effects caused by the presence of the topological edge channels[22] and is supported by the affinity of the $U_{xx}$-$I_{ds}$ and $U_{xy}$-$I_{ds}$ dependencies (see Figs. 2a and 2b). The frequency values derived from the $U_{xy}$-$I_{ds}$ characteristics are close to the above observations if the difference in length between the xx distance and the ds contact is considered.

Using the suggested technique, the comparison of the self-modes corresponding to the oligonucleotides of different lengths is rather intriguing. The resonant frequencies derived from the $dU_{xx}/dI_{ds}$ and $dU_{xy}/dI_{ds}$ dependencies on the $I_{ds}$ value appear to be significantly different for the 100- and 50-mer oligonucleotides (see Figs. 3a, 3b and 3c). This behaviour can be explained by the presence of a more complex relation between the self-resonant frequency and oligonucleotide lengths than provided by string resonance phenomena. Moreover, the data shown in Figs. 3a, 3b and 3c evidence about direct proportionality of the self-resonant frequency and the oligonucleotide length.

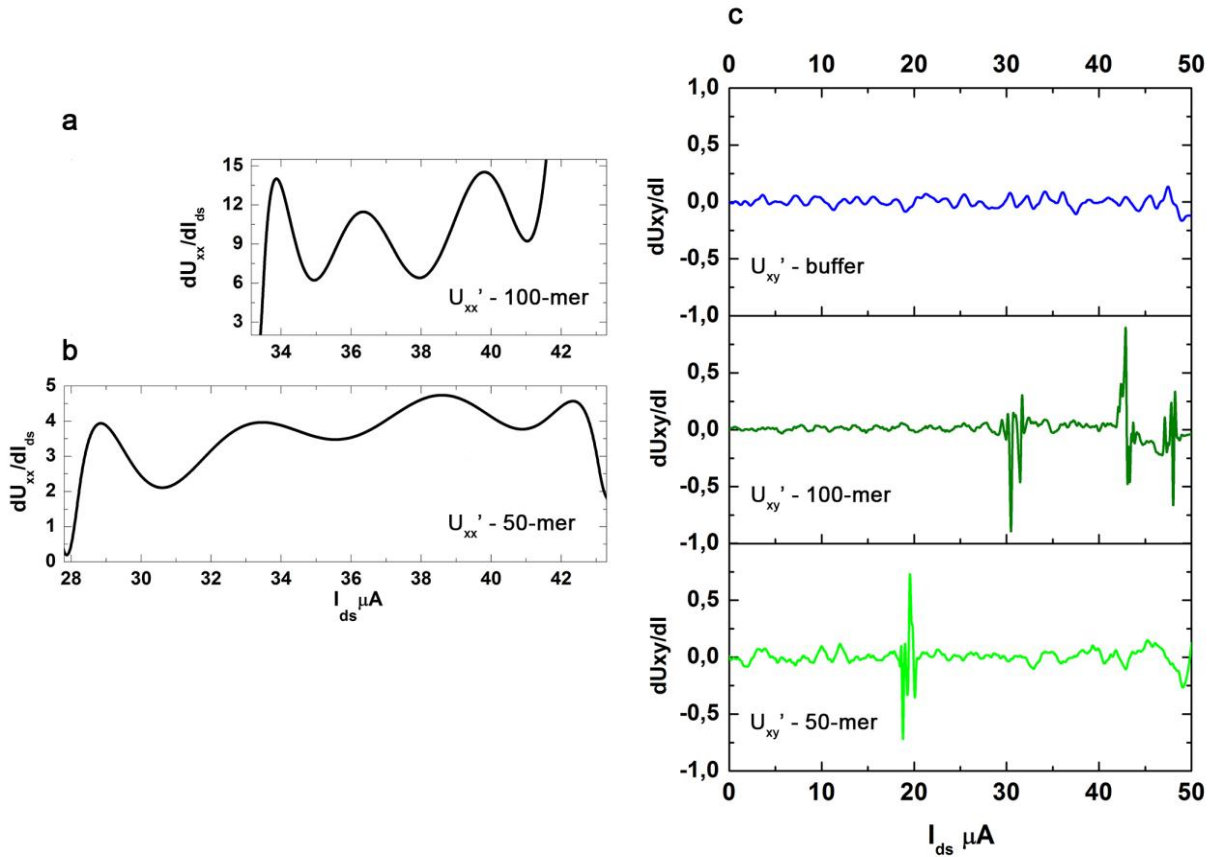

Figure 3 | **Derivatives of $U_{xx}$-$I_{ds}$ and $U_{xy}$-$I_{ds}$ characteristics.** The $U_{xx}$ derivatives corresponding to 100-mer (a) and 50-mer (b) oligonucleotides. c, The $U_{xy}$ derivatives corresponding to buffer solution (blue), 100-mer (dark green), and 50-mer (light green) oligonucleotides.

In this study, we demonstrate how a semiconductor nanostructure-based quantum pump device can be used as the sensing element for oligonucleotide detection systems. Experimental data in the section above correspond to the previous studies of oligonucleotides[11,12] and semiconductor nanostructures used in this research[18,23]. The quantum pump sensing techniques allow us to identify the self-modes of oligonucleotides. The resonant modes of the oligonucleotides were determined by analysing the $U_{xx}$-$I_{ds}$ and $U_{xy}$-$I_{ds}$ characteristics of the silicon nanosandwich with oligonucleotides deposited on its surface. All of the measurements were performed at room temperature in a sodium acetate buffer solution. The next step of this research is expected to provide a proper understanding of monomer and dimer molecule sensing. We anticipate that this novel approach will enable both the creation of new methods of label-free, real-time detection of nucleic acids and the modernisation of existing methods.

## METHODS SUMMARY
**Silicon nanosandwich preparation.** The devices are based on an ultra-narrow, 2 nm, high-mobility p-type Si-QW confined by delta barriers heavily doped with boron[24] ($5 \times 10^{21}$ cm$^{-3}$) on the n-type Si (100) surface[25]. These p-type Si-QWs are prepared on the n-type Si (100) wafers during the preliminary oxidation and subsequent short-time diffusion of boron by the CVD method[24,26,28,29]. The boron atoms have been shown to form trigonal dipole (B+ - B-) centres due to the negative-U reaction: $2B^o \rightarrow B^+ + B^-$ [26,27]. The conductance within the Si-QW is provided by the edge channels[30]. The sheet density of the holes has been determined by Hall-effect studies[18] to be $3 \times 10^{13}$ m$^{-2}$. Thus, there are 120 holes (*p*) in the edge channels. The system of microcavities has been formed on the surface of the silicon nanosandwich such that only a single hole and single oligonucleotide exist in one microcavity.

**Microfluidic system.** The container-type microfluidic system was constructed using polydimethylsiloxane. It is placed on the silicon nanosandwich surface and holds the 0.5 ml drop of solution, preventing it from evaporating. Each solution drop was deposited on the surface using a Proline plus Biohit 0.1-3 µl micropipette.

**Oligonucleotide preparation.** Single-stranded oligonucleotides were synthesised using Applied Biosystems synthesising equipment, purified with polyacrylamide gel electrophoresis, and extracted in a 0.3 M sodium acetate solution. The investigated sequences were 100 bp 5'-gcgctggctgcgggcggtgagctgagctcgcccccggggagctgtggccggcgcccctgccggttccctgagcagcggacgttcatgctgggagggcggcg-3' and 50 bp 5'-gcgctggctgcgggcggtgagctgagctcgcccccggggagctgtggccg-3'. The concentrations were 0.22 and 0.98 µg/µl, respectively. These concentrations were chosen to fit the concentration of holes in the edge channels.

**U-I measurement circuit.** The $U_{xx}$-$I_{ds}$ and $U_{xy}$-$I_{ds}$ measurement circuit consists of a DC source (Keithley 6221), $U_{xx}$ and $U_{xy}$ nanovoltmeters (Keithley 2182A), and the earthed metal capsule containing the chip holder. The system is synchronised using the National Instruments Lab View software package. The range of the driving DC drain-source current is (-50 – 50 µA), and the interval between the measured points is 100 nA. Each point was measured 10 times at a 1 ms interval.

1. Guiducci, C., Spiga, F. M. Another transistor-based revolution: on-chip qPCR. *Nat. Methods* **10**, 617-618 (2013).
2. Higuchi, R., Fockler, C., Dollinger, G., Watson, R. Kinetic PCR analysis: Real-Time monitoring of DNA Amplification reactions. *Nature Biotechnology* **11**, 1026 - 1030 (1993).
3. Udvardi, M.K., Czechowski, T., Scheible, W.R. Eleven Golden Rules of Quantitative RT-PCR. *The Plant Cell* **20**, 1736–1737 (2008).
4. VanGuilder, H.D., Vrana, K.E., Freeman, W.M. Twenty-five years of quantitative PCR for gene expression analysis. *BioTechniques* **44**, 619-626 (2008).
5. Liu, J., Liu, C., He, W. Fluorophores and Their Applications as Molecular Probes in Living Cells. *Current Organic Chemistry* **17**, 564-579 (2013)
6. Rothberg, J. M. *et al.* An integrated semiconductor device enabling non-optical genome sequencing. *Nature* **475**, 348-352 (2011).
7. Toumazou C. *et al.* Simultaneous DNA amplification and detection using a pH-sensing semiconductor system. *Nat. Methods* **10**, 641–646 (2013).
8. Kurz, V., Tanaka, T., Timp, G. Single Cell Transfection with Single Molecule Resolution Using a Synthetic Nanopore. *Nanoletters*, (2014).
9. Plesa, C., Ananth, A., Linko, V., Gülcher, C., Katan, A., Dietz, H., Dekker, C. Ionic permeability and mechanical properties of DNA origami nanoplates on solid-state nanopores. *ACS Nano*,(2014).
10. Venkatesan, B.M., Bashir, R. Nanopore sensors for nucleic acid analysis. *Nature Nanotechnology* **6**, (2011).
11. Thomas, G.J. Jr. Raman spectroscopy of protein and nucleic acid assemblies. *Annu. Rev. Biophys. Biomol. Struct.* **28**, 1–27 (1999).
12. Barhoumi, A., Zhang, D., Tam, F., Halas, N.G. Surface-Enhanced Raman Spectroscopy of DNA. *J. Am. Chem. Soc.* **130**, 5523–5529 (2008).
13. Webb, R. E. Field effect transistors for biological amplifiers. *Electronic Engrg.*, 803-805 (1965).
14. Lundstrom, I., Shlvaraman, M.S., Svenson, C.S., Lundkvlst, L. Hydrogen-sensitive MOS field effect transistor. *Appl Phys Lett* **26**, 55 – 57 (1975).
15. Bergveld, P. Thirty years of ISFETOLOGY What happened in the past 30 years and what may happen in the next 30 years. *Sensors and Actuators B* **88**, 1–20 (2003).

16. Lee, C.S., Kim, S.K., Kim, M. Ion-Sensitive Field-Effect Transistor for Biological Sensing. *Sensors* **9**, 7111-7131 (2009).
17. Bagraev, N.T., Klyachkin, L.E., Kudryavtsev, A.A., Malyarenko, A.M., Romanov, V.V. Superconductor Properties for Silicon Nanostructures. *Superconductor* **4**, 69-92 (2010).
18. Fischer, B.M., Walther, M., Uhd Jepsen, P. Far-infrared vibrational modes of DNA components studied by terahertz time-domain spectroscopy. *Phys. Med. Biol.* **47**, 3807(2002).
19. Blumenthal, M.D., Kaestner, B., Giblin, S., Janssen, T.J.B.M., Pepper, M., Anderson, D., Jones, G., Ritchie, A. Gigahertz quantized charge pumping. *Nature Physics* **3**, 343-347(2007).
20. Recher, P., Nazarov, Y.V., Kouwenhoven, L.P. Josephson Light-Emitting Diode. *PRL* **104**, 156802 (2010).
21. Josephson, B. D. The discovery of tunnelling supercurrents. *Rev. Mod. Phys.* **46**, 251 (1974).
22. Hasan M.Z., Kane C.L. Colloquium: Topological insulators. *Rev. Mod. Phys.* **82**, 3045 (2010).
23. Bagraev, N.T., Ivanov, V.K., Klyachkin, L.E., Shelykh, I.A. Spin depolarization in quantum wires polarize spontaneously in zero magnetic field. *Phys. Rev. B* **70**, 155315-155319 (2004).
24. Bagraev, N.T., Bouravleuv, A.D., Gehlhoff W, Klyachkin, L.E., Malyarenko, A.M., Rykov, S.A. Self-assembled impurity superlattices and microcavities in silicon. *Def. and Dif. Forum* **194**, 673(2001).
25. Bagraev, N.T., Galkin, N.G., Gehlhoff, W., Klyachkin, L.E., Malyarenko, A.M. Phase and amplitude response of the '0.7 feature' caused by holes in silicon one-dimensional wires and rings. *J. Phys.: Condens. Matter* **20**, 164202(2008).
26. Bagraev, N.T., Bouravleuv, A.D., Klyachkin, L.E., Malyarenko, A.M., Gehlhoff, W., Ivanov V.K., Shelykh, I.A. Quantum conductance in silicon quantum wires. *Semiconductors* **36**, 439 (2002).
27. Bagraev, N.T., Mashkov, V. A., Danilovskii, E. Yu., Gehlhoff, W., Gets, D.S., ., Klyachkin, L.E., Kudryavtsev, A.A., Kuzmin, R.V., Malyarenko, A.M., Romanov, V.V. EDESR and ODMR of impurity centers in nanostructures inserted in silicon microcavities. *Appl. Magn. Reson.* **39**, 113–135 (2010).
28. Bagraev, N.T., Bouravleuv, A.D., Klyachkin, L.E., Malyarenko, A.M., Gehlhoff, W., Romanov, Yu.I., Rykov, S.A. Local tunneling spectroscopy of silicon nanostructures. *Semiconductors* **39**, 716 (2005).
29. Bagraev, N.T., Gehlhoff, W., Klyachkin, L.E., Malyarenko, A.M., Romanov, V.V., Rykov, S.A. Superconductivity in silicon nanostructures. *Physica C* **219**, 437 (2006).
30. Bagraev, N.T., Danilovskii, E. Yu., Klyachkin, L.E., Malyarenko, A.M., Mashkov, V. A. Spin interference of holes in silicon nanosandwiches. *Semiconductors* **46**, 75-86 (2012).

**Acknowledgments** We are grateful to Dr A. M. Malyarenko, Dr D. S. Gets, and Dr E. Yu. Danilovskii for their helpful discussions and technical assistance. The reported study was mainly supported by the Russian Academy of Sciences and partially supported by the Russian Foundation for Basic Research in 2013 (Research Project No. 13-00-12055 ofi_m).